\documentclass[aps,prd,twocolumn,superscriptaddress]{revtex4}

\usepackage{mathtext}
\usepackage{amsmath}
\usepackage{amssymb}
\usepackage{latexsym}

\usepackage{graphicx}

\usepackage{epsfig}
\usepackage{epstopdf}

\begin{document}

\title{Relic neutrinos: Antineutrinos of Primordial Nucleosynthesis}


\author{Alexandre V. Ivanchik}\email{iav@astro.ioffe.ru}
 \affiliation{Ioffe Institute, Polytekhnicheskaya 26, St.-Petersburg 194021, Russia}

\author{Vlad Yu. Yurchenko}\email{yurchvlad.astro@mail.ioffe.ru}
 \affiliation{Ioffe Institute, Polytekhnicheskaya 26, St.-Petersburg 194021, Russia}

\date{\today}
\begin{abstract}
\vspace{2mm}	
For the first time the antineutrino spectrum formed as a result of neutron and tritium decays during the epoch  of primordial nucleosynthesis is calculated.
This spectrum is a non-thermal increase in addition to the standard cosmic neutrino background (C$\nu$B) whose thermal spectrum was formed before the beginning of primordial nucleosynthesis.
For energy larger than $10^{-2}$\,eV the calculated non-thermal antineutrino flux exceeds the C$\nu$B spectrum and there are no other comparable sources of antineutrino in this range.
The observations of these antineutrinos will allow us to look directly at the very early Universe and non-equilibrium processes taken place before, during, and some time after primordial nucleosynthesis.
\vspace{9mm}
\end{abstract}


\maketitle

\section{Introduction}

The observations of the Cosmic Microwave Background (CMB) allow us to see into our Universe when it was about 380\,000 years old.
Similarly primordial nucleosynthesis provides us with an indirect probe of the early Universe (about a few minutes old) based on the comparison of light element ($^4$He, D, $^7$Li) observations with corresponding theoretical calculations, which in turn is based on the well-established knowledge of nuclear and particle physics (e.g. \cite{Weinberg2008book,Gorbunov2011book}).  We cannot observe the Universe at that epoch directly using electromagnetic radiation due to the opacity of the Universe at early stages right up until primordial recombination. Nevertheless, the direct information about the first seconds of the Universe evolution principally can be obtained by the detection of relic neutrinos, well known as the cosmic neutrino background C$\nu$B (note that cosmic neutrino background consists of neutrino as well as antineutrino, so it is more correct to use the following abbreviation C$\nu\tilde{\nu}$B).
Like the cosmic microwave background radiation, the C$\nu$B was formed with a thermal equilibrium spectrum which for neutrinos $(\nu)$ and antineutrinos $(\tilde{\nu})$ is given by the Fermi-Dirac distribution (with zero neutrino mass):
 \begin{equation}
 n_{\nu,\tilde{\nu}}(p)dp = \frac{1}{(2\pi\hbar)^3}\frac{4\pi p^2 d p}{\exp(pc/kT)+1}
 \label{eq:F_D_dis}
 \end{equation}

Note that we can neglect the neutrino mass at the decoupling period ($T \sim 2$\,MeV) because there is the upper limit of $\sum m_{\nu}\!\!<\!\!0.23$\,eV \cite{Thomas2010, Ade2016AA} and the ratio $m_{\nu}c^2/kT$ is about $10^{-7}$ for this period.
After neutrino decoupling the spectrum has kept the same form (due to the adiabatic expansion of the Universe) with a temperature decreasing like   $T_{\nu}\!\propto\!(1+z)$, where $z$ is the cosmological redshift. It is very important to note that this fact takes place for momentum  distribution $n(p)$ whether or not neutrino possesses mass, while the form of energy distribution $n(\varepsilon)$ depends on neutrino mass (see next section). Therefore despite the fact that  $m_{\nu} \neq 0$, today $(z=0)$ the momentum distribution $n(p)$ has the form to be the same as Eq.\,(\ref{eq:F_D_dis}) with the current temperature $T_{\nu 0}$ whose value is related to the current temperature of the relic photons $T_{\gamma 0}$ (the CMB temperature). The theory of primordial nucleosynthesis give us the relation between the relic neutrino and photon temperatures, $T_{\nu}\!=\!(4/11)^{1/3}T_{\gamma}$, arising from electron-positron annihilation. Given this relation and the current value of $T_{\gamma 0}\!=\!2.725\pm0.001$\,K \cite{Fixsen2002ApJ}, we have the present temperature of relic neutrinos to be $T_{\nu 0}\!\approx\!1.945$\,K. However, at the same time the electron-positron annihilation leads to the heating of neutrinos and forms the non-thermal distortion of the neutrino spectrum \cite{Dolgov1992, Dodelson1992, Hannestad1995, Dolgov2002, Mangano2005}, therefore strictly speaking $T_{\nu 0}$ is not defined.

In addition to the relic neutrinos it should be mentioned that there are other neutrino backgrounds at high energies ($\varepsilon \gtrsim 1$\,MeV) which could be called ``cosmological'' as well: the diffuse supernova neutrino background \cite{Beacom2010}, the active galactic nuclei (AGN) background and cosmogenic (GZK) neutrinos arising from interactions of ultra-energetic protons with the CMB photons (for more details see e.g. \cite{Katz2012} and references therein). 

In this work we discuss the complementary ability to look at the early Universe and non-equilibrium processes occurred before, during and after primordial nucleosynthesis using spectrum calculations and future possible observations of antineutrinos having arisen from neutron and triton (nucleus of tritium) decays (${\rm n} \rightarrow {\rm p} + e^{-} + \tilde{\nu}$, $\text{t} \rightarrow {\rm ^3he} + e^{-} + \tilde{\nu}$).

\section{The Spectrum of Antineutrino of Primordial Nucleosynthesis}

\begin{figure}
\includegraphics[width=\columnwidth]{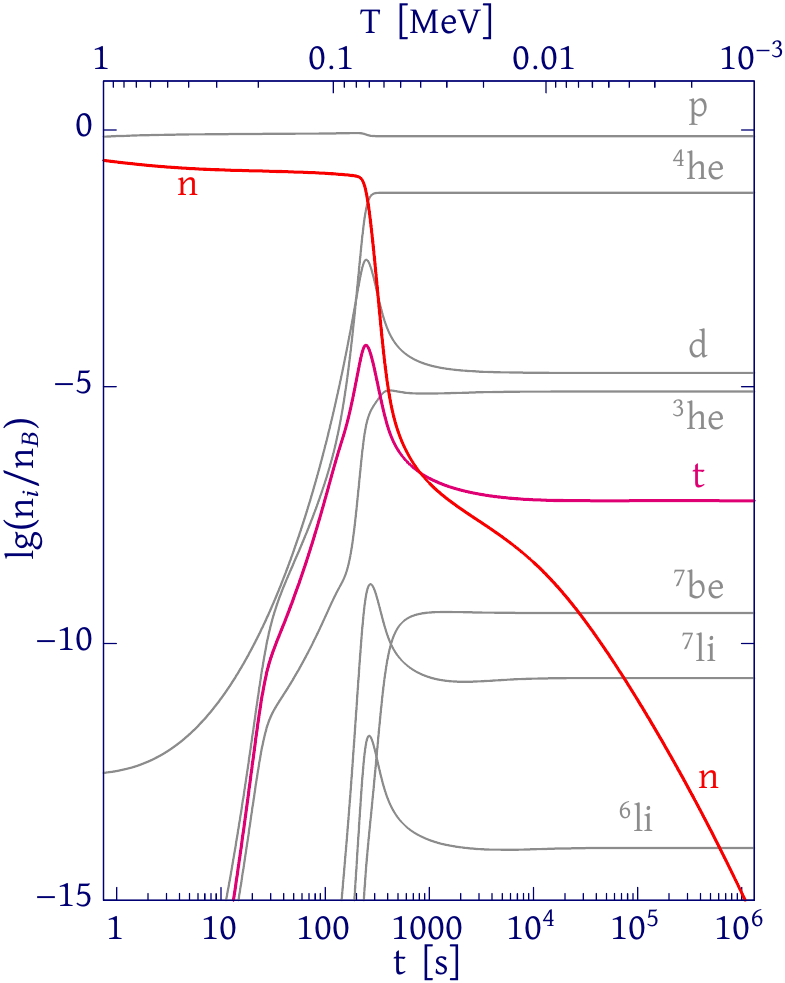}
\protect\caption{\label{fig:bbn4} The time-temperature evolution of the relative abundances $Y_x\equiv n_x/n_b$ of light nuclei produced during  primordial nucleosynthesis. Red and crimson lines correspond to neutron and tritium relative abundances. The calculations are done by using our own numerical code of primordial nucleosynthesis \cite{Orlov2000AAT} (see the text).}
\end{figure}

During primordial nucleosynthesis as a result of weak and nuclear interactions there is some amount of neutron and tritium (see Fig.\,\ref{fig:bbn4}), whose decays give rise to non-thermal antineutrinos in addition to the thermal relic neutrinos and antineutrinos (Eq.\,\ref{eq:F_D_dis}).

The number densities of antineutrinos arising under neutron and tritium decays with momentum between $p$ and $p+dp$ during the time $dt$
is given by
 \begin{equation}
  dn_{\nu}(p) = \lambda n_{n,t}(t) f(p) dp dt
 \label{eq:Nu_spec_t}
 \end{equation}
Here $\lambda$ is neutron or tritium decay rate. It relates with the mean lifetime as $\lambda\!\!=\!\!1/\tau$ ($\tau_n\!\!=\!\!(880.2\pm1.0)$\,s \cite{Patrignani2016} and $\tau_t\!\!=\!\!(17.66\!\pm\!0.03)$\,yr \cite{Aku}),
$n_{n,t}(t)$ is the number density of neutron/tritium at the moment $t$ corresponding cosmological epoch with redshift $z$.
It can be expressed as $n_{n,t}(z)\!=\!Y_{n,t}(z)n_b(z)\!=\!Y_{n,t}(z)n^0_b(1\!+\!z)^3$, where $Y_{n,t}(z)$ is the ratio of neutrons or tritium to all baryons $n_b$, and $n^0_b\!=\!\eta n^0_{\gamma}\simeq2.48\cdot10^{-7}\,$cm$^{-3}$ (see, e.g. \cite{Olive_2014PDG}) is the baryon number density at the present epoch $(z\!=\!0)$.  The value $f(p)dp$ is the neutrino fraction with momentum between
$p$ and $p+dp$ arising due to neutron/tritium decays
%
 \begin{equation}
 f(p) d p =  \mathcal{C} \sqrt{(1-\tilde{p})(1-\tilde{p}+2\tilde{m}_e)}(1-\tilde{p}+\tilde{m}_e) \tilde{p}^2 d\tilde{p}
 \label{eq:Beta_decay_profile}
 \end{equation}
%
here, neutrino momentum takes a value $0\leqslant p \leqslant Q/c$,  $Q$~is the value of the energy release of the reaction, $\tilde{p}=pc/Q$ is dimensionless
momentum, $0\leqslant \tilde{p} \leqslant 1$, $\tilde m_e=m_ec^2/Q$ -- dimensionless electron mass, $\mathcal{C}$ is the dimensionless normalization constant determined by
%
 \begin{equation*}
 \int_0^{Q/c}\!\!f(p)dp=1.
 \label{eq:Beta_decay_profile_normalization}
 \end{equation*}

Since the spectrum has kept its form under the adiabatic expansion of the Universe and neutrino that has momentum $p_0$ at the present epoch $(z=0)$ had momentum $p(z)=p_0\!\cdot\!(1+z)$ at cosmological epoch with redshift $z$, therefore we can express the modern momentum distribution in the following form:
%
 \begin{equation}
 f_0(p_0)dp_0=f(p_0\cdot\!(1{+}z))(1{+}z)dp_0 \, ,
 \label{eq:Beta_decay_profile_redshifted}
 \end{equation}
%
where $f(p_0\cdot\!(1{+}z))$ is taken from the right-hand side of Eq.\,(\ref{eq:Beta_decay_profile}). 

Therefore, the spectrum of the non-thermal antineutrinos formed at the BBN era and observed now can be expressed as

 \begin{equation}
  \frac{dn_{\nu}(p_0)}{dp_0} = -\frac{1}{\tau}\!\!\int\limits_0^{\tilde{z}(p_0)}\!\!\frac{n(z)}{(1+z)^3}f(p_0(1+z))
  (1+z)\frac{dt}{dz} dz,
 \label{eq:Nu_spec}
 \end{equation}
here the minus sign reflects the fact that redshift decreased as time passed;
$p_0$ takes a value from the range $0\leqslant p_0 \leqslant Q/c$, for given $p_0$ the value $p_0(1+z)$ cannot be large than $Q/c$ what gives us the upper limit of integration $\tilde{z}(p_0)=Q/(p_0c)-1$, at the same time this value must not be larger than redshift of neutrino decoupling; the factor $1/(1+z)^3$ in the integral  arises from the dilution of neutrons and tritium due to the Universe expansion. The function $dt/dz$ in Eq.\,(\ref{eq:Nu_spec}) can be derived from the time-temperature-redshift relation (e.g. \cite{Weinberg2008book}):
\begin{equation}
t= 1.78\,\,\mbox{sec}\cdot\!\left[ \frac{T_{\gamma0} (1+z)}{10^{10}\mbox{K}} \right]^{-2}\!\!\!\!\!\!\! + \text{Const}
\end{equation}
which is valid for this cosmological period, when the early Universe was radiation dominated. To obtain dependences of the relative abundances $n(z)$ of light nuclei on redshift during the period under consideration, we have updated our own previous numerical code of primordial nucleosynthesis \cite{Orlov2000AAT} which is based on the historical Wagoner's code \cite{Wagoner1967}.
Results obtained by using our code (Fig.\,\ref{fig:bbn4}) are in satisfactory agreement with the results of modern analogous codes \citep{Smith1993,Arbey2011,Pisanti2008,Pitrou2018}.

\begin{figure*}
	\includegraphics[width=2\columnwidth]{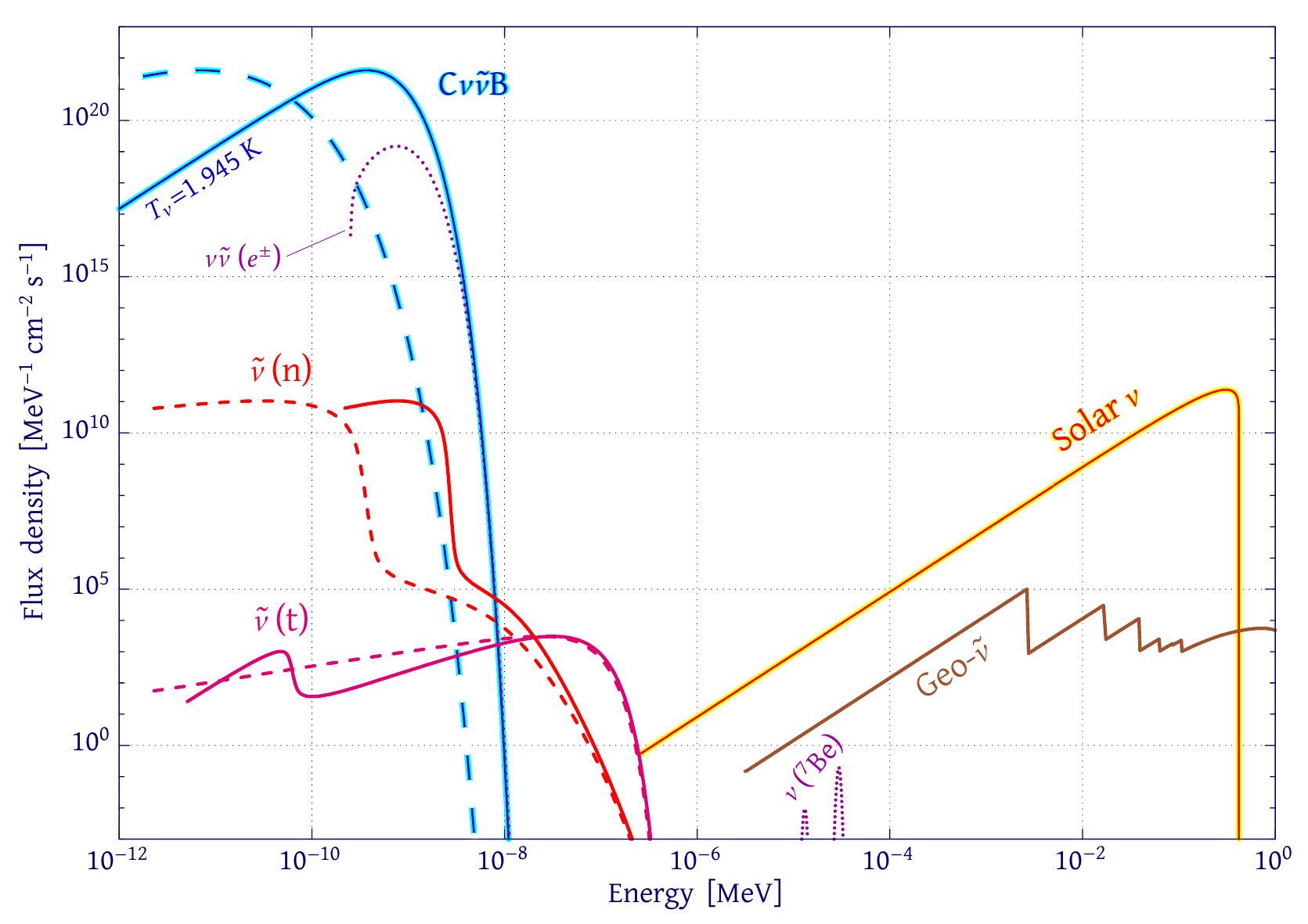}
	\protect\caption{\label{fig:res_pr_antinu}
		{\bf Neutrino and antineutrino spectra}. The relic neutrino and antineutrino equilibrium background C$\nu\tilde{\nu}$B ($\equiv$C$\nu$B) is determined by Eq.\ref{eq:F_D_dis} (blue-cyan solid curve for $m_{\nu}=0$). The solar neutrino spectrum (red-yellow curve) constructed by using data from \citep{Bahcall1988}. The geologically produced antineutrino spectrum (brown curve) \cite{Enomoto2006,Enomoto2007}.  {\bf The non-thermal spectra}. Doublet of narrow cosmological neutrino lines due to decays of $^7$Be  \citep{Khatri2011} and the non-equilibrium neutrino distortion $(\Delta f \equiv f-f_0)$ of the C$\nu$B high-energy tail caused by $e^{\pm}$-annihilation \cite{Dolgov1992, Dodelson1992} (dotted violet curves). The non-thermal antineutrino spectra formed by decays of neutrons $\tilde{\nu}({\rm n})$ (red curve) and tritons $\tilde{\nu}({\rm t})$ (crimson curve) which have been calculated in this work. In addition to the spectra mentioned above we present the spectra of  C$\nu$B and  $\tilde{\nu}({\rm n})$, $\tilde{\nu}({\rm t})$ calculated under conditions of non-zero neutrino mass (dashed curves; for instance, we take $m_{\nu}=0.01$\,eV, and the energy for this case denotes $\varepsilon_{kin} \equiv \varepsilon -m_{\nu}c^2$).}
\end{figure*}

The calculated antineutrino spectra, together with the C$\nu$B neutrino/antineutrino, the Solar neutrino, and the geologically produced antineutrino spectra are shown in Fig.\,\ref{fig:res_pr_antinu}. It should be noted that the calculated antineutrino spectra for both neutrons and tritium have specific features. The neutron spectrum has a sharp step and the tritium spectrum shows a local maximum. These features repeat the existing peculiarities of the time evolution of the neutron and tritium abundances (Fig.\,\ref{fig:bbn4}). 

In the case of massless neutrinos with the relativistic energy-momentum relation $\varepsilon=pc$ the neutrino energy distribution function and the neutrino momentum distribution function (Eq.\,\ref{eq:Nu_spec}) are the same up to the constant factor $d\varepsilon/dp=c$, while in the case $m_{\nu}\neq 0$, with the energy-momentum relation $\varepsilon=\sqrt{p^2c^2+m^2c^4}$, it has led to significant differences of the distributions in the kinetic energy range where $\varepsilon_{kin} \equiv \varepsilon -mc^2 < m_{\nu}c^2$ (see Fig.\,\ref{fig:res_pr_antinu}).      

Since a flux of particles is experimentally observable quantity, this is what we show on Fig.\,\ref{fig:res_pr_antinu}. Flux of antineutrino can be found by integrating  Eq.\,\eqref{eq:Nu_spec} multiplied by a neutrino velocity $v(\varepsilon)=d\varepsilon(p)/dp$ over hemisphere in the projection on normal to field on which flux is measured:
%
 \begin{equation}
 \frac{dF(\varepsilon)}{d\varepsilon}=\frac{1}{4\pi}\int\!\frac{dn_{\nu}(\varepsilon)}{d\varepsilon}v(\varepsilon)\cos\Theta d\Omega.
 \label{eq:spectral_flux}
 \end{equation}
%
As spatial distribution of antineutrino is isotropic integrating Eq.\,\eqref{eq:spectral_flux} is reduced to multiplying of Eq.\,\eqref{eq:Nu_spec} by a factor $v(\varepsilon)/4$. In addition, it is important to note that, in comparing the total fluxes of solar and cosmological neutrinos moving through the surface of an imaginary neutrino detector, solar neutrino would penetrate into the detector from only one hemisphere while cosmological neutrino and antineutrino would penetrate into the interior of the detector through both hemispheres.

\vspace{-4mm}
\section{Three sorts of non-thermal relic neutrinos.}
\vspace{-4mm}
Thus, there are three sorts of non-thermal relic neutrinos related to different physical phenomena of primordial nucleosynthesis:
\begin{itemize}
	\item The first of these phenomena is the non-equilibrium neutrino distortion of the C$\nu$B high-energy tail caused by $e^{\pm}$-annihilation occurred at the beginning of primordial nucleosynthesis at $T\lesssim 1$\,MeV  (for more details see \cite{Dolgov1992, Dodelson1992, Hannestad1995, Dolgov2002, Mangano2005}).
	\item The second one is narrow neutrino lines from $^7$Be decay happened after primordial nucleosynthesis and before primordial hydrogen recombination \cite{Khatri2011}. Note that the authors briefly mentioned about the existence of non-thermal antineutrinos from n and t decays but they did not calculate their spectra.    
	\item Finally, the non-thermal antineutrinos from n and t decays whose spectra were calculated in this work for the first time. 
\end{itemize}

For comparison all three non-thermal relic neutrino spectra are shown in Fig.\,\ref{fig:res_pr_antinu}. Unfortunately, the first two mentioned spectra have the very small contrast against the background of the C$\nu$B and solar neutrino spectra, while n and t decay antineutrino spectra considerably exceed the C$\nu$B and any known antineutrino background for energy larger than $10^{-2}$\,eV.     

\section{Possibility of the direct detection of the relic neutrinos}

The existence of C$\nu$B has been indirectly proved from the very good agreement between primordial nucleosynthesis predictions and astronomical observations of the light element abundances, $^4$He, D, $^7$Li (e.g. \cite{Cyburt2016}).

The ideas of detecting  C$\nu$B have been discussed since the 1960s. However the direct observations of the relic neutrinos is a great challenge to present experimental techniques due to the very low energy $(\lesssim 10^{-4}$\,eV) of relic neutrinos at the present epoch. A principal possibility of the direct detection of the relic neutrino background was first discussed in papers \cite{Weinberg1962} and \cite{Irvine1983JourPhysG}. Among several proposals for the direct detection of the relic neutrinos, the most promising ones are relic neutrino capture on radioactive $\beta$-decaying nuclei \cite{Cocco2007JCAP} and relic antineutrinos capture on radioactive nuclei decaying via electron capture \cite{Cocco2009PhysRevD, Lusignoli2011PhysLettB}. The descriptions of other ways and current perspectives of the direct detection of the relic neutrinos/antineutrinos see, for instance, the  following detailed works \cite{Ringwald2009NuPhA, PTOLEMY, Li2015IJMPA, Faessler2015JPhCS, Birrell2015EPJC, XingZhou2011}.

\section{Conclusion}

The non-thermal relic neutrinos provide a powerful tool to probe the early Universe.
The CMB discovery became definitely an important milestone in cosmology. First, we have learned that the unique physical phenomenon, an echo of early hot stages of the Universe, exists. Second, the study of the CMB distortions gives us information about non-equilibrium processes which have been taking place from primordial recombination up to now, and provides accurate values of several key cosmological parameters \cite{Ade2016AA}. Likewise, the observations of the C$\nu$B spectrum and its distortions will become a unique window on the much more early Universe (a few minutes and hours after the Big Bang). \\

{\em Acknowledgments.} The authors thank the anonymous referee
for valuable comments and suggestions. The work has been supported by Russian Science Foundation (Grant 18-12-00301).

\bibliography{pnn4}

\begin{thebibliography}{37}
\expandafter\ifx\csname natexlab\endcsname\relax\def\natexlab#1{#1}\fi
\expandafter\ifx\csname bibnamefont\endcsname\relax
  \def\bibnamefont#1{#1}\fi
\expandafter\ifx\csname bibfnamefont\endcsname\relax
  \def\bibfnamefont#1{#1}\fi
\expandafter\ifx\csname citenamefont\endcsname\relax
  \def\citenamefont#1{#1}\fi
\expandafter\ifx\csname url\endcsname\relax
  \def\url#1{\texttt{#1}}\fi
\expandafter\ifx\csname urlprefix\endcsname\relax\def\urlprefix{URL }\fi
\providecommand{\bibinfo}[2]{#2}
\providecommand{\eprint}[2][]{\url{#2}}

\bibitem[{\citenamefont{{Weinberg}}(2008)}]{Weinberg2008book}
\bibinfo{author}{\bibfnamefont{S.}~\bibnamefont{{Weinberg}}},
  \emph{\bibinfo{title}{{Cosmology}}} (\bibinfo{publisher}{Oxford University
  Press}, \bibinfo{year}{2008}).

\bibitem[{\citenamefont{{Gorbunov} and {Rubakov}}(2011)}]{Gorbunov2011book}
\bibinfo{author}{\bibfnamefont{D.~S.} \bibnamefont{{Gorbunov}}}
  \bibnamefont{and} \bibinfo{author}{\bibfnamefont{V.~A.}
  \bibnamefont{{Rubakov}}}, \emph{\bibinfo{title}{{Introduction to the Theory
  of the Early Universe: Hot Big Bang Theory}}} (\bibinfo{publisher}{World
  Scientific Publishing Co}, \bibinfo{year}{2011}).

\bibitem[{\citenamefont{{Thomas} et~al.}(2010)\citenamefont{{Thomas},
  {Abdalla}, and {Lahav}}}]{Thomas2010}
\bibinfo{author}{\bibfnamefont{S.~A.} \bibnamefont{{Thomas}}},
  \bibinfo{author}{\bibfnamefont{F.~B.} \bibnamefont{{Abdalla}}},
  \bibnamefont{and} \bibinfo{author}{\bibfnamefont{O.}~\bibnamefont{{Lahav}}},
  \bibinfo{journal}{Physical Review Letters} \textbf{\bibinfo{volume}{105}},
  \bibinfo{eid}{031301} (\bibinfo{year}{2010}), \eprint{0911.5291}.

\bibitem[{\citenamefont{{Planck Collaboration}
  et~al.}(2016)\citenamefont{{Planck Collaboration}, {Ade}, {Aghanim},
  {Arnaud}, {Ashdown}, {Aumont}, {Baccigalupi}, {Banday}, {Barreiro},
  {Bartlett} et~al.}}]{Ade2016AA}
\bibinfo{author}{\bibnamefont{{Planck Collaboration}}},
  \bibinfo{author}{\bibfnamefont{P.~A.~R.} \bibnamefont{{Ade}}},
  \bibinfo{author}{\bibfnamefont{N.}~\bibnamefont{{Aghanim}}},
  \bibinfo{author}{\bibfnamefont{M.}~\bibnamefont{{Arnaud}}},
  \bibinfo{author}{\bibfnamefont{M.}~\bibnamefont{{Ashdown}}},
  \bibinfo{author}{\bibfnamefont{J.}~\bibnamefont{{Aumont}}},
  \bibinfo{author}{\bibfnamefont{C.}~\bibnamefont{{Baccigalupi}}},
  \bibinfo{author}{\bibfnamefont{A.~J.} \bibnamefont{{Banday}}},
  \bibinfo{author}{\bibfnamefont{R.~B.} \bibnamefont{{Barreiro}}},
  \bibinfo{author}{\bibfnamefont{J.~G.} \bibnamefont{{Bartlett}}},
  \bibnamefont{et~al.}, \bibinfo{journal}{Astronomy \& Astrophysics}
  \textbf{\bibinfo{volume}{594}}, \bibinfo{eid}{A13} (\bibinfo{year}{2016}),
  \eprint{1502.01589}.

\bibitem[{\citenamefont{{Fixsen} and {Mather}}(2002)}]{Fixsen2002ApJ}
\bibinfo{author}{\bibfnamefont{D.~J.} \bibnamefont{{Fixsen}}} \bibnamefont{and}
  \bibinfo{author}{\bibfnamefont{J.~C.} \bibnamefont{{Mather}}},
  \bibinfo{journal}{The Astrophysical Journal} \textbf{\bibinfo{volume}{581}},
  \bibinfo{pages}{817} (\bibinfo{year}{2002}).

\bibitem[{\citenamefont{{Dolgov} and {Fukugita}}(1992)}]{Dolgov1992}
\bibinfo{author}{\bibfnamefont{A.~D.} \bibnamefont{{Dolgov}}} \bibnamefont{and}
  \bibinfo{author}{\bibfnamefont{M.}~\bibnamefont{{Fukugita}}},
  \bibinfo{journal}{Soviet Journal of Experimental and Theoretical Physics
  Letters} \textbf{\bibinfo{volume}{56}}, \bibinfo{pages}{123}
  (\bibinfo{year}{1992}).

\bibitem[{\citenamefont{{Dodelson} and {Turner}}(1992)}]{Dodelson1992}
\bibinfo{author}{\bibfnamefont{S.}~\bibnamefont{{Dodelson}}} \bibnamefont{and}
  \bibinfo{author}{\bibfnamefont{M.~S.} \bibnamefont{{Turner}}},
  \bibinfo{journal}{\prd} \textbf{\bibinfo{volume}{46}}, \bibinfo{pages}{3372}
  (\bibinfo{year}{1992}).

\bibitem[{\citenamefont{{Hannestad} and {Madsen}}(1995)}]{Hannestad1995}
\bibinfo{author}{\bibfnamefont{S.}~\bibnamefont{{Hannestad}}} \bibnamefont{and}
  \bibinfo{author}{\bibfnamefont{J.}~\bibnamefont{{Madsen}}},
  \bibinfo{journal}{\prd} \textbf{\bibinfo{volume}{52}}, \bibinfo{pages}{1764}
  (\bibinfo{year}{1995}), \eprint{astro-ph/9506015}.

\bibitem[{\citenamefont{{Dolgov}}(2002)}]{Dolgov2002}
\bibinfo{author}{\bibfnamefont{A.~D.} \bibnamefont{{Dolgov}}},
  \bibinfo{journal}{Physics Reports} \textbf{\bibinfo{volume}{370}},
  \bibinfo{pages}{333} (\bibinfo{year}{2002}), \eprint{hep-ph/0202122}.

\bibitem[{\citenamefont{{Mangano} et~al.}(2005)\citenamefont{{Mangano},
  {Miele}, {Pastor}, {Pinto}, {Pisanti}, and {Serpico}}}]{Mangano2005}
\bibinfo{author}{\bibfnamefont{G.}~\bibnamefont{{Mangano}}},
  \bibinfo{author}{\bibfnamefont{G.}~\bibnamefont{{Miele}}},
  \bibinfo{author}{\bibfnamefont{S.}~\bibnamefont{{Pastor}}},
  \bibinfo{author}{\bibfnamefont{T.}~\bibnamefont{{Pinto}}},
  \bibinfo{author}{\bibfnamefont{O.}~\bibnamefont{{Pisanti}}},
  \bibnamefont{and} \bibinfo{author}{\bibfnamefont{P.~D.}
  \bibnamefont{{Serpico}}}, \bibinfo{journal}{Nuclear Physics B}
  \textbf{\bibinfo{volume}{729}}, \bibinfo{pages}{221} (\bibinfo{year}{2005}),
  \eprint{hep-ph/0506164}.

\bibitem[{\citenamefont{{Beacom}}(2010)}]{Beacom2010}
\bibinfo{author}{\bibfnamefont{J.~F.} \bibnamefont{{Beacom}}},
  \bibinfo{journal}{Annual Review of Nuclear and Particle Science}
  \textbf{\bibinfo{volume}{60}}, \bibinfo{pages}{439} (\bibinfo{year}{2010}),
  \eprint{1004.3311}.

\bibitem[{\citenamefont{{Katz} and {Spiering}}(2012)}]{Katz2012}
\bibinfo{author}{\bibfnamefont{U.~F.} \bibnamefont{{Katz}}} \bibnamefont{and}
  \bibinfo{author}{\bibfnamefont{C.}~\bibnamefont{{Spiering}}},
  \bibinfo{journal}{Progress in Particle and Nuclear Physics}
  \textbf{\bibinfo{volume}{67}}, \bibinfo{pages}{651} (\bibinfo{year}{2012}),
  \eprint{1111.0507}.

\bibitem[{\citenamefont{{Orlov} et~al.}(2000)\citenamefont{{Orlov}, {Ivanchik},
  and {Varshalovich}}}]{Orlov2000AAT}
\bibinfo{author}{\bibfnamefont{A.~V.} \bibnamefont{{Orlov}}},
  \bibinfo{author}{\bibfnamefont{A.~V.} \bibnamefont{{Ivanchik}}},
  \bibnamefont{and} \bibinfo{author}{\bibfnamefont{D.~A.}
  \bibnamefont{{Varshalovich}}}, \bibinfo{journal}{Astronomical and
  Astrophysical Transactions} \textbf{\bibinfo{volume}{19}},
  \bibinfo{pages}{375} (\bibinfo{year}{2000}).

\bibitem[{\citenamefont{Patrignani et~al.}(2016)}]{Patrignani2016}
\bibinfo{author}{\bibfnamefont{C.}~\bibnamefont{Patrignani}}
  \bibnamefont{et~al.} (\bibinfo{collaboration}{Particle Data Group}),
  \bibinfo{journal}{Chin. Phys.} \textbf{\bibinfo{volume}{C40}},
  \bibinfo{pages}{100001} (\bibinfo{year}{2016}).

\bibitem[{\citenamefont{{Akulov} and {Mamyrin}}(2005)}]{Aku}
\bibinfo{author}{\bibfnamefont{Y.~A.} \bibnamefont{{Akulov}}} \bibnamefont{and}
  \bibinfo{author}{\bibfnamefont{B.~A.} \bibnamefont{{Mamyrin}}},
  \bibinfo{journal}{Phys. Lett. B} \textbf{\bibinfo{volume}{610}},
  \bibinfo{pages}{45} (\bibinfo{year}{2005}).

\bibitem[{\citenamefont{{Olive} and {Particle Data
  Group}}(2014)}]{Olive_2014PDG}
\bibinfo{author}{\bibfnamefont{K.~A.} \bibnamefont{{Olive}}} \bibnamefont{and}
  \bibinfo{author}{\bibnamefont{{Particle Data Group}}},
  \bibinfo{journal}{Chinese Physics C} \textbf{\bibinfo{volume}{38}},
  \bibinfo{eid}{090001} (\bibinfo{year}{2014}).

\bibitem[{\citenamefont{{Wagoner} et~al.}(1967)\citenamefont{{Wagoner},
  {Fowler}, and {Hoyle}}}]{Wagoner1967}
\bibinfo{author}{\bibfnamefont{R.~V.} \bibnamefont{{Wagoner}}},
  \bibinfo{author}{\bibfnamefont{W.~A.} \bibnamefont{{Fowler}}},
  \bibnamefont{and} \bibinfo{author}{\bibfnamefont{F.}~\bibnamefont{{Hoyle}}},
  \bibinfo{journal}{\apj} \textbf{\bibinfo{volume}{148}}, \bibinfo{pages}{3}
  (\bibinfo{year}{1967}).

\bibitem[{\citenamefont{{Smith} et~al.}(1993)\citenamefont{{Smith}, {Kawano},
  and {Malaney}}}]{Smith1993}
\bibinfo{author}{\bibfnamefont{M.~S.} \bibnamefont{{Smith}}},
  \bibinfo{author}{\bibfnamefont{L.~H.} \bibnamefont{{Kawano}}},
  \bibnamefont{and} \bibinfo{author}{\bibfnamefont{R.~A.}
  \bibnamefont{{Malaney}}}, \bibinfo{journal}{Astrophysical Journal Supplement
  Series} \textbf{\bibinfo{volume}{85}}, \bibinfo{pages}{219}
  (\bibinfo{year}{1993}).

\bibitem[{\citenamefont{{Arbey}}(2012)}]{Arbey2011}
\bibinfo{author}{\bibfnamefont{A.}~\bibnamefont{{Arbey}}},
  \bibinfo{journal}{Computer Physics Communications}
  \textbf{\bibinfo{volume}{183}}, \bibinfo{pages}{1822} (\bibinfo{year}{2012}),
  \eprint{1106.1363}.

\bibitem[{\citenamefont{{Pisanti} et~al.}(2008)\citenamefont{{Pisanti},
  {Cirillo}, {Esposito}, {Iocco}, {Mangano}, {Miele}, and
  {Serpico}}}]{Pisanti2008}
\bibinfo{author}{\bibfnamefont{O.}~\bibnamefont{{Pisanti}}},
  \bibinfo{author}{\bibfnamefont{A.}~\bibnamefont{{Cirillo}}},
  \bibinfo{author}{\bibfnamefont{S.}~\bibnamefont{{Esposito}}},
  \bibinfo{author}{\bibfnamefont{F.}~\bibnamefont{{Iocco}}},
  \bibinfo{author}{\bibfnamefont{G.}~\bibnamefont{{Mangano}}},
  \bibinfo{author}{\bibfnamefont{G.}~\bibnamefont{{Miele}}}, \bibnamefont{and}
  \bibinfo{author}{\bibfnamefont{P.~D.} \bibnamefont{{Serpico}}},
  \bibinfo{journal}{Computer Physics Communications}
  \textbf{\bibinfo{volume}{178}}, \bibinfo{pages}{956} (\bibinfo{year}{2008}),
  \eprint{0705.0290}.

\bibitem[{\citenamefont{{Pitrou} et~al.}(2018)\citenamefont{{Pitrou}, {Coc},
  {Uzan}, and {Vangioni}}}]{Pitrou2018}
\bibinfo{author}{\bibfnamefont{C.}~\bibnamefont{{Pitrou}}},
  \bibinfo{author}{\bibfnamefont{A.}~\bibnamefont{{Coc}}},
  \bibinfo{author}{\bibfnamefont{J.-P.} \bibnamefont{{Uzan}}},
  \bibnamefont{and}
  \bibinfo{author}{\bibfnamefont{E.}~\bibnamefont{{Vangioni}}},
  \bibinfo{journal}{ArXiv e-prints} \bibinfo{eid}{arXiv:1801.08023}
  (\bibinfo{year}{2018}).

\bibitem[{\citenamefont{{Bahcall} and {Ulrich}}(1988)}]{Bahcall1988}
\bibinfo{author}{\bibfnamefont{J.~N.} \bibnamefont{{Bahcall}}}
  \bibnamefont{and} \bibinfo{author}{\bibfnamefont{R.~K.}
  \bibnamefont{{Ulrich}}}, \bibinfo{journal}{Reviews of Modern Physics}
  \textbf{\bibinfo{volume}{60}}, \bibinfo{pages}{297} (\bibinfo{year}{1988}).

\bibitem[{\citenamefont{{Enomoto}}(2006)}]{Enomoto2006}
\bibinfo{author}{\bibfnamefont{S.}~\bibnamefont{{Enomoto}}},
  \bibinfo{journal}{Earth Moon and Planets} \textbf{\bibinfo{volume}{99}},
  \bibinfo{pages}{131} (\bibinfo{year}{2006}).

\bibitem[{\citenamefont{{Enomoto} et~al.}(2007)\citenamefont{{Enomoto},
  {Ohtani}, {Inoue}, and {Suzuki}}}]{Enomoto2007}
\bibinfo{author}{\bibfnamefont{S.}~\bibnamefont{{Enomoto}}},
  \bibinfo{author}{\bibfnamefont{E.}~\bibnamefont{{Ohtani}}},
  \bibinfo{author}{\bibfnamefont{K.}~\bibnamefont{{Inoue}}}, \bibnamefont{and}
  \bibinfo{author}{\bibfnamefont{A.}~\bibnamefont{{Suzuki}}},
  \bibinfo{journal}{Earth and Planetary Science Letters}
  \textbf{\bibinfo{volume}{258}}, \bibinfo{pages}{147} (\bibinfo{year}{2007}).

\bibitem[{\citenamefont{{Khatri} and {Sunyaev}}(2011)}]{Khatri2011}
\bibinfo{author}{\bibfnamefont{R.}~\bibnamefont{{Khatri}}} \bibnamefont{and}
  \bibinfo{author}{\bibfnamefont{R.~A.} \bibnamefont{{Sunyaev}}},
  \bibinfo{journal}{Astronomy Letters} \textbf{\bibinfo{volume}{37}},
  \bibinfo{pages}{367} (\bibinfo{year}{2011}), \eprint{1009.3932}.

\bibitem[{\citenamefont{{Cyburt} et~al.}(2016)\citenamefont{{Cyburt}, {Fields},
  {Olive}, and {Yeh}}}]{Cyburt2016}
\bibinfo{author}{\bibfnamefont{R.~H.} \bibnamefont{{Cyburt}}},
  \bibinfo{author}{\bibfnamefont{B.~D.} \bibnamefont{{Fields}}},
  \bibinfo{author}{\bibfnamefont{K.~A.} \bibnamefont{{Olive}}},
  \bibnamefont{and} \bibinfo{author}{\bibfnamefont{T.-H.} \bibnamefont{{Yeh}}},
  \bibinfo{journal}{Reviews of Modern Physics} \textbf{\bibinfo{volume}{88}},
  \bibinfo{eid}{015004} (\bibinfo{year}{2016}), \eprint{1505.01076}.

\bibitem[{\citenamefont{{Weinberg}}(1962)}]{Weinberg1962}
\bibinfo{author}{\bibfnamefont{S.}~\bibnamefont{{Weinberg}}},
  \bibinfo{journal}{Physical Review} \textbf{\bibinfo{volume}{128}},
  \bibinfo{pages}{1457} (\bibinfo{year}{1962}).

\bibitem[{\citenamefont{{Irvine} and {Humphreys}}(1983)}]{Irvine1983JourPhysG}
\bibinfo{author}{\bibfnamefont{J.~M.} \bibnamefont{{Irvine}}} \bibnamefont{and}
  \bibinfo{author}{\bibfnamefont{R.}~\bibnamefont{{Humphreys}}},
  \bibinfo{journal}{Journal of Physics G: Nuclear Physics}
  \textbf{\bibinfo{volume}{9}}, \bibinfo{pages}{847} (\bibinfo{year}{1983}).

\bibitem[{\citenamefont{{Cocco} et~al.}(2007)\citenamefont{{Cocco}, {Mangano},
  and {Messina}}}]{Cocco2007JCAP}
\bibinfo{author}{\bibfnamefont{A.~G.} \bibnamefont{{Cocco}}},
  \bibinfo{author}{\bibfnamefont{G.}~\bibnamefont{{Mangano}}},
  \bibnamefont{and}
  \bibinfo{author}{\bibfnamefont{M.}~\bibnamefont{{Messina}}},
  \bibinfo{journal}{Journal of Cosmology and Astroparticle Physics}
  \textbf{\bibinfo{volume}{6}}, \bibinfo{eid}{015} (\bibinfo{year}{2007}).

\bibitem[{\citenamefont{{Cocco} et~al.}(2009)\citenamefont{{Cocco}, {Mangano},
  and {Messina}}}]{Cocco2009PhysRevD}
\bibinfo{author}{\bibfnamefont{A.~G.} \bibnamefont{{Cocco}}},
  \bibinfo{author}{\bibfnamefont{G.}~\bibnamefont{{Mangano}}},
  \bibnamefont{and}
  \bibinfo{author}{\bibfnamefont{M.}~\bibnamefont{{Messina}}},
  \bibinfo{journal}{Physical Review D} \textbf{\bibinfo{volume}{79}},
  \bibinfo{pages}{053009} (\bibinfo{year}{2009}).

\bibitem[{\citenamefont{{Lusignoli} and
  {Vignati}}(2011)}]{Lusignoli2011PhysLettB}
\bibinfo{author}{\bibfnamefont{M.}~\bibnamefont{{Lusignoli}}} \bibnamefont{and}
  \bibinfo{author}{\bibfnamefont{M.}~\bibnamefont{{Vignati}}},
  \bibinfo{journal}{Physics Letters B} \textbf{\bibinfo{volume}{697}},
  \bibinfo{pages}{11} (\bibinfo{year}{2011}), \eprint{1012.0760}.

\bibitem[{\citenamefont{{Ringwald}}(2009)}]{Ringwald2009NuPhA}
\bibinfo{author}{\bibfnamefont{A.}~\bibnamefont{{Ringwald}}},
  \bibinfo{journal}{Nuclear Physics A} \textbf{\bibinfo{volume}{827}},
  \bibinfo{pages}{501} (\bibinfo{year}{2009}), \eprint{0901.1529}.

\bibitem[{\citenamefont{{Betts et al}}(2013)}]{PTOLEMY}
\bibinfo{author}{\bibfnamefont{S.}~\bibnamefont{{Betts et al}}},
  \bibinfo{journal}{ArXiv e-prints} \bibinfo{eid}{arXiv:1307.4738}
  (\bibinfo{year}{2013}).

\bibitem[{\citenamefont{{Li}}(2015)}]{Li2015IJMPA}
\bibinfo{author}{\bibfnamefont{Y.-F.} \bibnamefont{{Li}}},
  \bibinfo{journal}{International Journal of Modern Physics A}
  \textbf{\bibinfo{volume}{30}}, \bibinfo{eid}{1530031} (\bibinfo{year}{2015}),
  \eprint{1504.03966}.

\bibitem[{\citenamefont{{Faessler} et~al.}(2015)\citenamefont{{Faessler},
  {Hodak}, {Kovalenko}, and {Simkovic}}}]{Faessler2015JPhCS}
\bibinfo{author}{\bibfnamefont{A.}~\bibnamefont{{Faessler}}},
  \bibinfo{author}{\bibfnamefont{R.}~\bibnamefont{{Hodak}}},
  \bibinfo{author}{\bibfnamefont{S.}~\bibnamefont{{Kovalenko}}},
  \bibnamefont{and}
  \bibinfo{author}{\bibfnamefont{F.}~\bibnamefont{{Simkovic}}},
  \bibinfo{journal}{Journal of Physics Conference Series}
  \textbf{\bibinfo{volume}{580}}, \bibinfo{eid}{012040} (\bibinfo{year}{2015}).

\bibitem[{\citenamefont{{Birrell} and {Rafelski}}(2015)}]{Birrell2015EPJC}
\bibinfo{author}{\bibfnamefont{J.}~\bibnamefont{{Birrell}}} \bibnamefont{and}
  \bibinfo{author}{\bibfnamefont{J.}~\bibnamefont{{Rafelski}}},
  \bibinfo{journal}{European Physical Journal C} \textbf{\bibinfo{volume}{75}},
  \bibinfo{eid}{91} (\bibinfo{year}{2015}), \eprint{1402.3409}.

\bibitem[{\citenamefont{{Xing} and {Zhou}}(2011)}]{XingZhou2011}
\bibinfo{author}{\bibfnamefont{Z.-Z.} \bibnamefont{{Xing}}} \bibnamefont{and}
  \bibinfo{author}{\bibfnamefont{S.}~\bibnamefont{{Zhou}}},
  \emph{\bibinfo{title}{Neutrinos in Particle Physics, Astronomy and
  Cosmology}} (\bibinfo{publisher}{Springer-Verlag Berlin Heidelberg},
  \bibinfo{year}{2011}).

\end{thebibliography}

\end{document}